\documentclass[twocolumn,english,aps,prb,showpacs,superscriptaddress]{revtex4}
\usepackage[T1]{fontenc}
\usepackage[latin9]{inputenc}
\usepackage{graphicx,graphics,epsfig}
\usepackage{babel}

\begin{document}

\title{Exchange engineering in 3d chains adsorbed  on Cu$_2$N/Cu(001)}
\author{M.C. Urdaniz}
\affiliation{Departamento de F\'{\i}sica, Comisi\'on Nacional de Energ\'{\i}a At\'omica,
Gral. Paz 1499, 1650 San Mart\'{\i}n, Buenos Aires, Argentina.}
\author{M.A. Barral}
\affiliation{Departamento de F\'{\i}sica, Comisi\'on Nacional de Energ\'{\i}a At\'omica,
Gral. Paz 1499, 1650 San Mart\'{\i}n, Buenos Aires, Argentina.}
\author{A.M. Llois}
\affiliation{Departamento de F\'{\i}sica, Comisi\'on Nacional de Energ\'{\i}a At\'omica,
Gral. Paz 1499, 1650 San Mart\'{\i}n, Buenos Aires, Argentina.}
\date{\today }

\begin{abstract}

Covalent substrates  can give rise to a  variety of  magnetic interaction mechanisms among 
adsorbed transition metal atoms 
building atomic nanostructures. We show this by calculating the ground state magnetic configuration 
of monoatomic $3d$ chains deposited on a  monolayer of  Cu$_2$N  grown on Cu(001) as a function of 
$d$ filling and of adsorption sites of the one dimensional nanostructures.
\end{abstract}

\maketitle

\section{Introduction}

The study of  the  interactions which underlie the magnetic ground state 
of deposited  nanostructures  constitutes an area of enormous interest nowadays, both, in 
applied and in  basic research.  The systems of interest can be  atomically 
manipulated and deposited  on different substrates by means of  
scanning tunneling microcopy (STM). For future  spintronic applications 
the search is, among others,  towards adequate films able to decouple the deposited nanostructures 
from the underlying metallic substrate, diminishing as much as possible the influence  
of the latter on the electronic and magnetic properties of  adsorbed   nano-objects. 
Actually, isolation is never complete and,  the decoupling  layers themselves play, in principle,  
a role to be determined on the  magnetic  interactions of the nanostructures.

Copper nitride overlayers grown on different  Cu surfaces  have  been recently 
used as  decoupling systems, being it possible to develop  stable thin films even at  
the monolayer thickness, while preserving an ordered geometric structure.  
Actually, the number of systems which fulfill these conditions  seems to 
be limited and,  thereafter, copper nitride monolayers, called from now 
on Cu$_2$N,  are currently being  investigated both  theoretically as 
experimentally.\cite{Stampfl, Gupta1, Gupta2} 

Focusing on the magnetic interactions within the nanostructures, Mn chains 
of different lengths have been deposited  by Hirjibehedin 
\emph{et al}~\cite{Hirjibehedin} on Cu$_2$N/Cu(001) 
at different adsorption sites by doing manipulation with an STM. 
Performing  inelastic electron tunneling spectroscopy (IETS) 
the authors concluded that  the  interatomic  
Mn-Mn interactions are antiferromagnetic with an interaction strength which 
depends on the deposition configuration.  There have been different 
theoretical contributions addressing these 
interactions.\cite{jones,nuestro1,nuestro2,Jones_Ti,Jones_Lin,Scopel} In particular, 
in previous works we have studied  the effect of the  surrounding geometry 
on the magnetic interatomic interactions for Mn and Cr chains deposited on 
Cu$_2$N/Cu(001).\cite{nuestro1,nuestro2}

Recently the group of R. Wiesendanger has proposed to make use 
of the RKKY interaction among Fe adatoms deposited on a non magnetic metallic 
susbstrate, namely, the (111) surface of a copper monocrystal, to fabricate spin-based 
logic devices.\cite{wiesendanger}  In fact, this group has achieved it to realize 
and develop a model system  able to 
perform logic operations. The tip of an STM was used to  deposit 
chains of antiferromagnetically 
coupled Fe adatoms and, using spin polarized 
scanning-tunneling spectroscopy (SP-STS), the magnetic 
interaction among the spins of the deposited  adatoms was measured.
The designed  model device  was able to  transmit and process 
certain desired  magnetic 
information. The authors suggest that combinations  of antiferromagnetic 
and ferromagnetically coupled chains could be used to realize new 
model gates by 
properly tuning the interatomic distances of the adsorbed atoms. 

In this contribution we show that using covalent decoupling layers, 
such as the Cu$_2$N monolayers grown  on Cu(001), it is possible to 
obtain different intrachain magnetic interactions for an  adsorbed $3d$ 
atomic chain with a given 
$d$ filling.  This can be achieved not by tuning  the  interatomic distance of the adsorbed
nanostructures but, by changing the adsorption sites of the transition metal atoms,
which could be part of an eventual spintronic device. The richness provided by  the 
molecular substrate lies in the fact that the interatomic  interactions within  the deposited  
chains can change dramatically not only as a funtion of $d$-filling  but also depending 
on adsorption  configuration, while keeping the distance among chain  atoms  fixed.

We present the results of  a  systematic study of  the behaviour  of the 
magnetic interactions within Cr, Mn, Fe and Co chains deposited on the 
above  mentioned molecular system. Four  different  
chain adsorption geometries are considered, geometries   which  involve  
different magnetic mediation processes,  with the aim of  highlighting 
the effect of the local and crystal  
environment on the interacting  $3d$ orbitals of the chains.  We find 
that the adsorption sites of the chain atoms on the decoupling layer 
determine the kind of magnetic 
interaction involved, while the filling  of the  $d$-orbitals determines the final magnetic ground state.

\section{Calculation Details}

We perform \emph{ab initio} calculations based on density functional 
theory (DFT) using the full
potential linearized augmented plane waves method (FP-LAPW), as 
implemented in the WIEN2k code.\cite{wien}
The generalized gradient approximation (GGA) for the exchange and correlation 
potential in the parametrization of Perdew, Burke and Ernzerhof~\cite{perdew}
and the augmented plane wave-local orbital (apw-lo) basis are used.
The cutoff parameter which gives 
the number of plane waves in the interstitial region is taken as 
$R_{mt} K_{max}=7$, where $K_{max}$ is the value of the largest reciprocal 
lattice vector used in the
plane waves' expansion and $R_{mt}$ is the smallest \emph{muffin tin} 
radio used. The number of $\vec{k}$ points in the Brillouin zone is enough in 
each case to obtain the desired energy and charge
precision, namely  $10^{-4}$ Ry  and  $10^{-4}$ e respectively.

The nitrogen atoms of the pristine substrate occupy fourfold coordinated hollow sites 
on the outermost metallic copper layer, 
forming a c(2x2) local structure.\cite{Leibsle}
This structure forms islands on the Cu(001) surface and, as mentioned in
the introduction, the chains are built atom by atom  on one of these Cu$_2$N islands.
We simulate an island by considering a Cu$_2$N monolayer on Cu(001). This
system is modeled with a supercell composed by a five layers slab separated 
by a vacuum region of 13.8 \AA. 
The thickness of the vacuum region is found to be 
sufficiently large  to avoid interactions
among subsequent slabs. To exploit  inversion symmetry and to prevent
unphysical multipoles,  each slab has one Cu$_2$N monolayer on each side.

The surface lattice constant follows pseudomorphically the underlying Cu(001) substrate
and we use the experimental lattice  constant of copper. We consider that the  adsorbed
3$d$ atomic chains are infinite long and  they are deposited on both sides of the slabs.
The nearest neighbour 3$d$-3$d$  distance along the chains is  3.61 \AA  \ in all the cases studied.  
In these  supercell calculations,  chains are periodically  arranged 7.2 \AA  \ apart from each other
assuming  that the interchain interactions are much smaller than the intrachain ones. 

The positions of all atoms in the supercell of 
the proposed  systems are allowed to move until forces are smaller 
than 0.1 eV/\AA.

\section{Calculations and Results}

\subsection{The different surface structures}

\begin{figure*}[ht]
\centering
\begin{tabular}{p{3.5cm}p{3.5cm}p{3.5cm}p{3.5cm}p{3.5cm}}
\epsfig{file=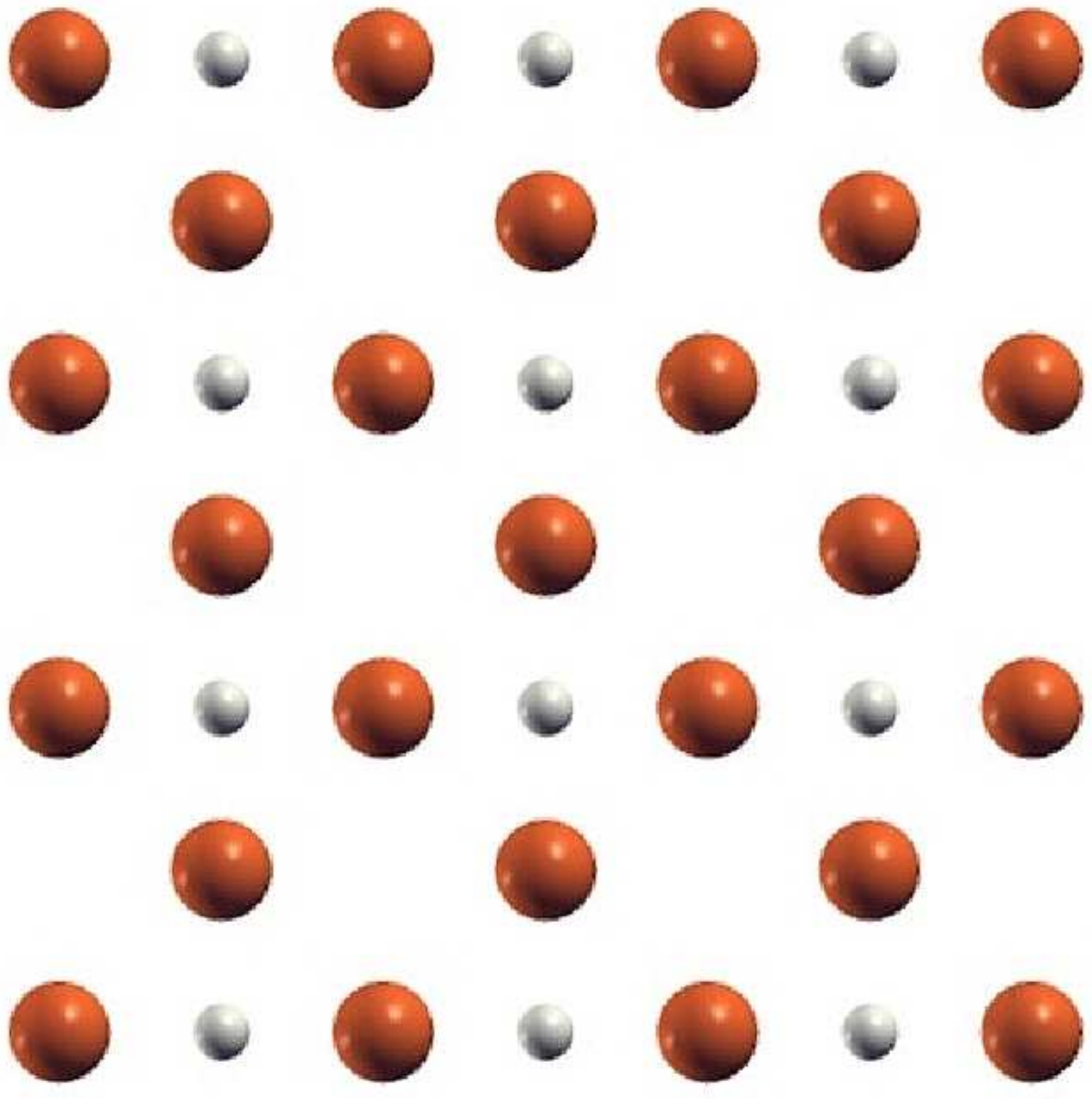,width=3.2cm}   &
\epsfig{file=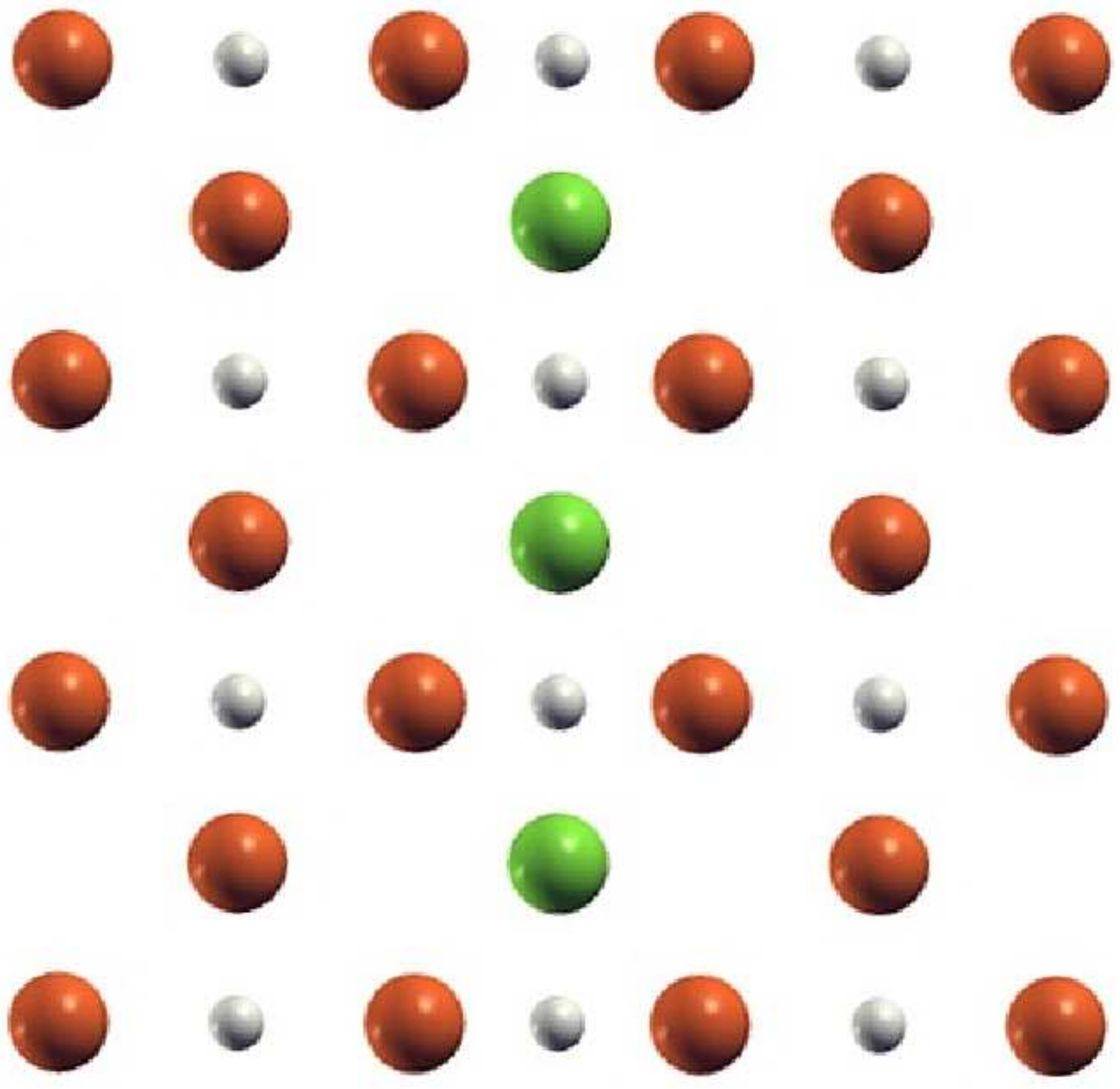,width=3.2cm} &
\epsfig{file=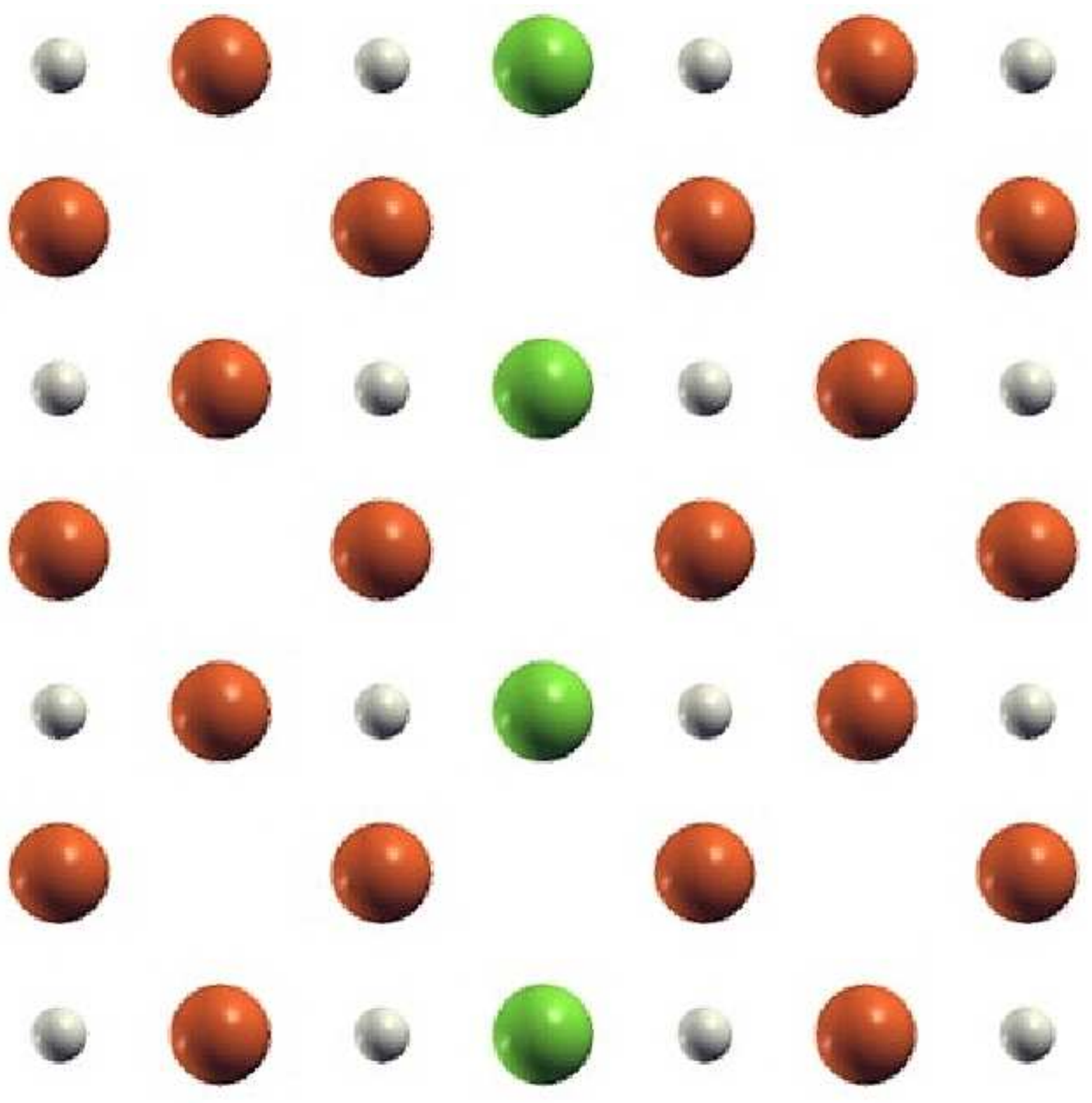,width=3.2cm} &
\epsfig{file=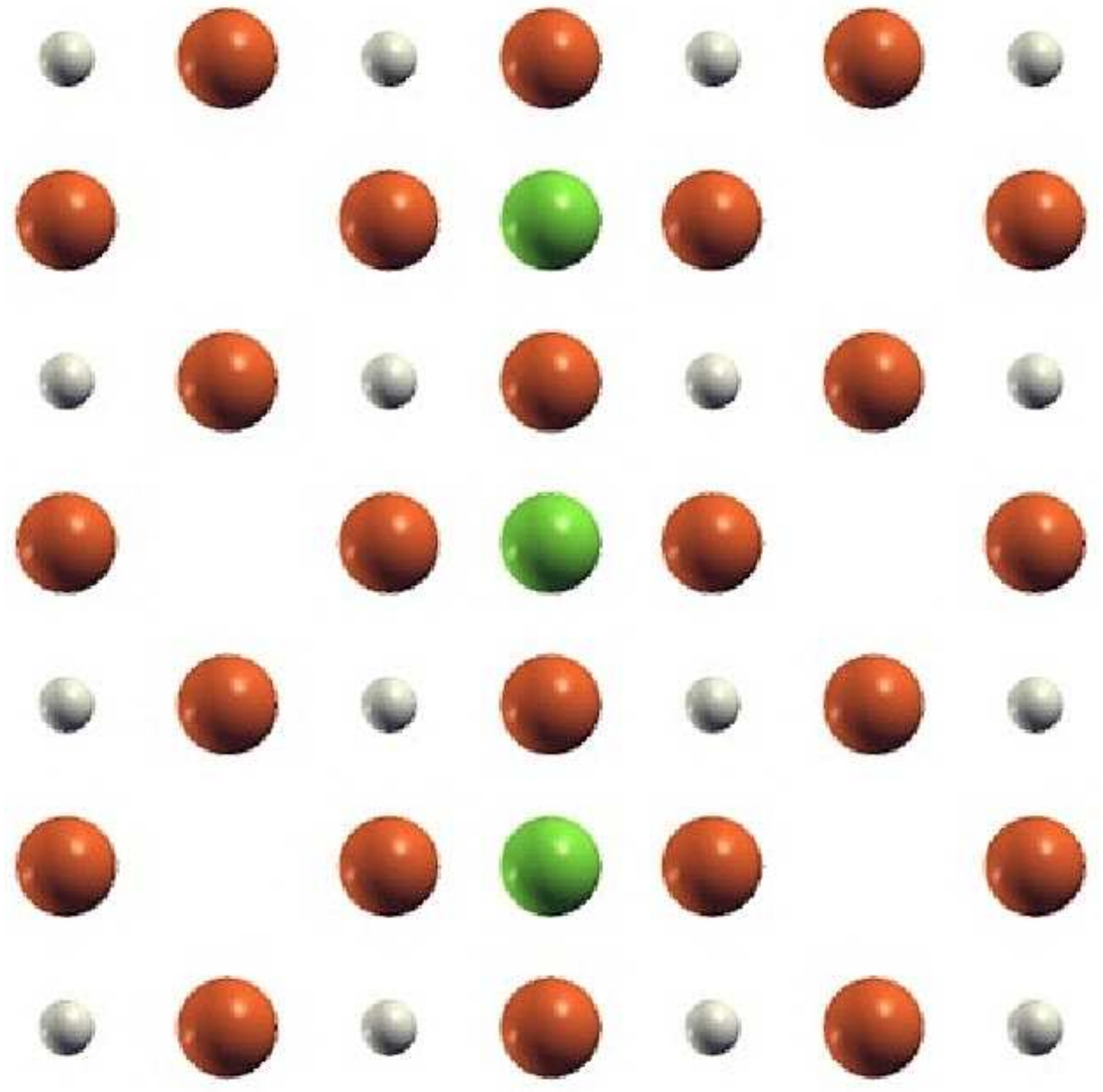,width=3.2cm} &
\epsfig{file=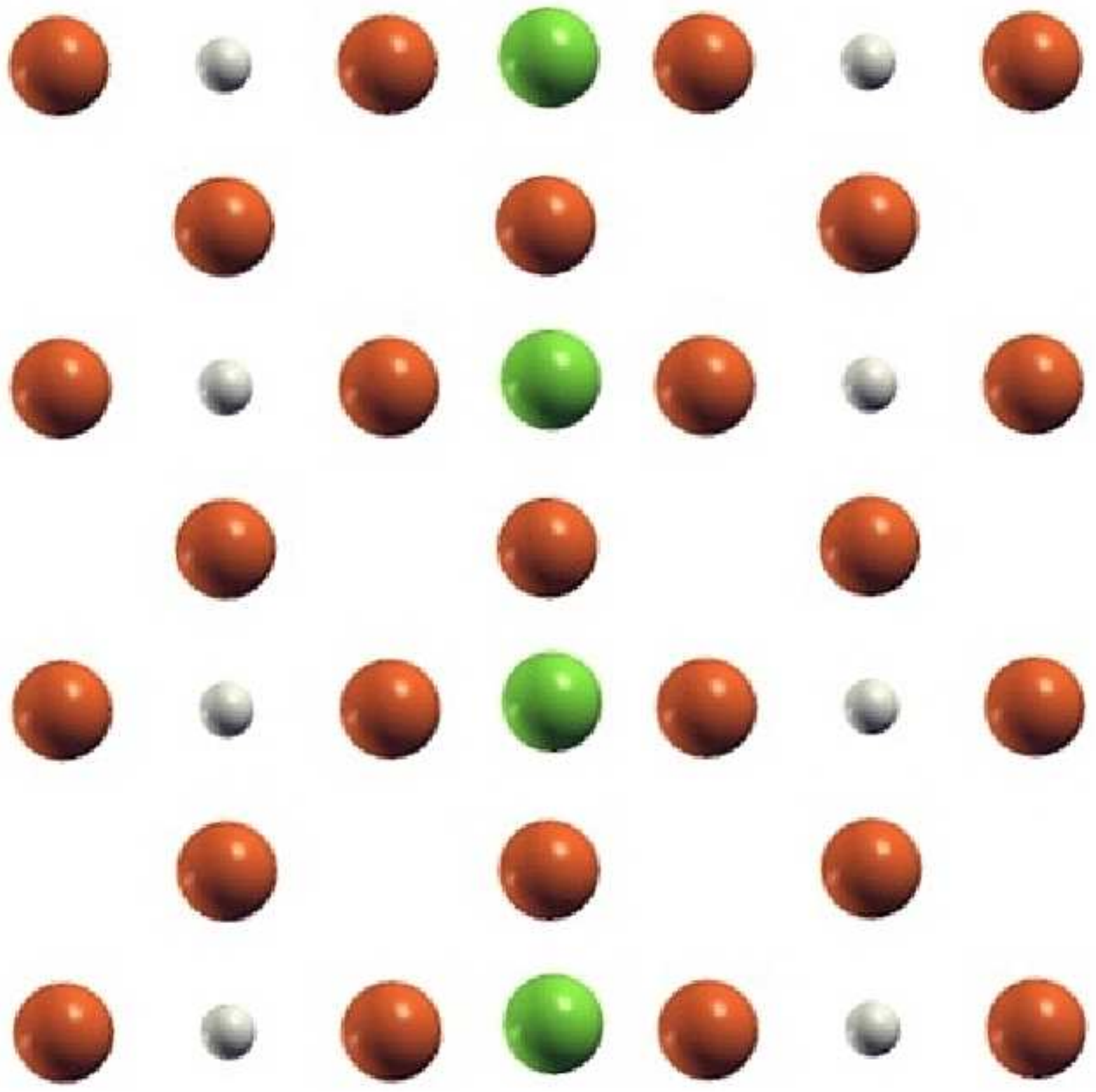,width=3.2cm}    \\
\hspace{13mm}( a) & \hspace{13mm} (b) &\hspace{13mm} (c) &\hspace{13mm} (d) &\hspace{13mm} (e)    \\
\end{tabular}
\caption{\label{fig:fig3} (Color online) Schematic top view of the different
adsorption configurations of the 3$d$ chains on Cu$_{2}$N/Cu(001) considered.
(a) Pristine Cu$_{2}$N  monolayer, (b)-(e) correspond to the \emph{s1},
\emph{s2}, \emph{s3} and \emph{s4} arrangements.
TM, N and Cu atoms are depicted by green, white and red spheres respectively.}
\end{figure*}

For Cr, Mn, Fe and Co chains, we consider four different stable 
adsorption geometries. Fig.~\ref{fig:fig3} (a) displays a
schematic top view of the Cu$_2$N/Cu(001) substrate, while the different 
geometric chain configurations  are depicted in Fig.~\ref{fig:fig3} (b-e).

In the first adsorption geometry, the 3$d$ chains are arranged directly on top of Cu 
atoms of the Cu$_2$N monolayer,  in such a way that one N atom of the 
nitride lies inbetween each pair of $3d$ atoms along the chain.
There are four lateral  copper atoms of the Cu$_2$N monolayer  
surrounding each chain atom.
This is one of the experimentally 
reported situations\cite{Hirjibehedin} and we call it \emph{s1}. See 
Fig.~\ref{fig:fig3}(b).

In the second geometry, \emph{s2}, each atom of the chain is deposited, 
as in the case of \emph{s1},  on top of Cu atoms but now every chain 
atom is laterally 
surrounded by two nitrogens,  one at each side in the direction perpendicular 
to the chain, as shown in Fig.~\ref{fig:fig3} (c). There   are also  
four lateral copper atoms of the Cu$_2$N substrate 
surrounding each chain atom. 

In the third case, called the \emph{s3} arrangement, the 3$d$ 
atoms are located in hollow sites of the Cu$_2$N layer,  as it is shown
in Fig.~\ref{fig:fig3}(d). There are four surface Cu atoms and four 
nitrogens around each 3$d$ one.

The last configuration considered, the \emph{s4},  is one in which the 
3$d$ atoms sit on top of nitrogens of the substrate as depicted in  
Fig.~\ref{fig:fig3} (e). Each 3$d$ atom is surrounded by four
coppers of the underlying substrate.  
This configuration has also been experimentally reported. \cite{Hirjibehedin}

Due to the reduced coordination number and the covalent nature of the underlying Cu$_2$N 
network, the local  atomic environment should play,   a priori,  a relevant   
role on the final  electronic  and magnetic structure   of the  deposited chains, being it  important to
properly optimize them. We relax  the different systems by   starting always from the ferromagnetic spin
configuration of the  chain arrangement.

For the different 3$d$ fillings the final relaxed positions 
within  each configuration are very similar, the differences among them 
being at most of 0.1 \AA.
In the \emph{s1} geometry  the N atoms, which are inbetween 
two 3$d$ ones, change significantly their vertical distance  with respect to the Cu$_2$N 
layer, relaxing outwards and forming  a nearly linear diatomic chain 
together with the 3$d$ atoms. 
These last ones lie slightly higher (0.1 \AA) than the N atoms,  the bond
length between them is around 1.8 \AA \ for all $d$ fillings considered.
This is in agreement with previous results.\cite{jones}
The Cu atoms just below  the diatomic chain are second nearest neighbors of the 
3$d$ ones.  The distance between them lies in the range 2.3-2.4 \AA.  There are
four  Cu atoms on the surface layer which are third nearest neighbours 
to a 3$d$ one  at 2.7-2.8 \AA \ distance.

When relaxing the \emph{s2} structure, 
the two lateral N atoms of each 3$d$ one  move towards the chains 
forming N-3$d$ 
bonds  of lengths lying  in the range 1.8-1.9 \AA. 
The distance between the  3$d$  transition
metal atoms and the Cu atom directly beneath them ends up being,  
after relaxation, of  2.3-2.4 \AA.
There are  four lateral copper atoms at 2.9 \AA \ from each 3$d$ one.

The effect of relaxing the atomic positions in the \emph{s3} configuration is
that each 3$d$ atom ends up having, in the Cu$_2$N monolayer, four copper 
atoms at 2.4-2.6 \AA \ and four nitrogens  at  2.5 \AA \ . 
The copper atom beneath each 3$d$ one, which belongs to the metallic Cu(001) 
substrate is at 3.4-3.5 \AA \ of the 
corresponding chain atom.

In the case of the \emph{s4} structure the distance between each 3$d$
chain atom and the nitrogen sitting directly below, ranges 
from 1.7 \AA \ to  1.8 \AA \ depending on $d$ filling. In this 
case the N atoms 
move away from the Cu$_2$N monolayer towards the $3d$ chain and 
the four nearest copper atoms are at 3.0-3.1 \AA \ distance from each 
chain atom. 

In all the adsorption geometries considered, the N atoms closest to the 
$3d$ chains have a tendency to relax towards the transition metal chain atoms.

\subsection{Bonding and electronic structure.}

The electronic density distribution of the systems under study 
depends on the adsorption  sites of the chains. It determines
the type of bonding and, as it will be seen in the next section, 
also the nature of the magnetic  interactions. 

Analyzing the electronic density distribution, we find that
there are three different ways in which the considered adsorbed  3$d$ 
chains bind to the substrate: i) forming a diatomic  
covalent one dimensional nanostructure  "metalically" bonded to the 
substrate, ii) building a monoatomic chain which binds covalently to
the substrate and, iii) building a monoatomic  chain which binds
metalically to the substrate.

\begin{figure}[ht]
\centering
\begin{tabular}{p{3.8cm}p{3.8cm}p{0.6cm}}
\psfig{file=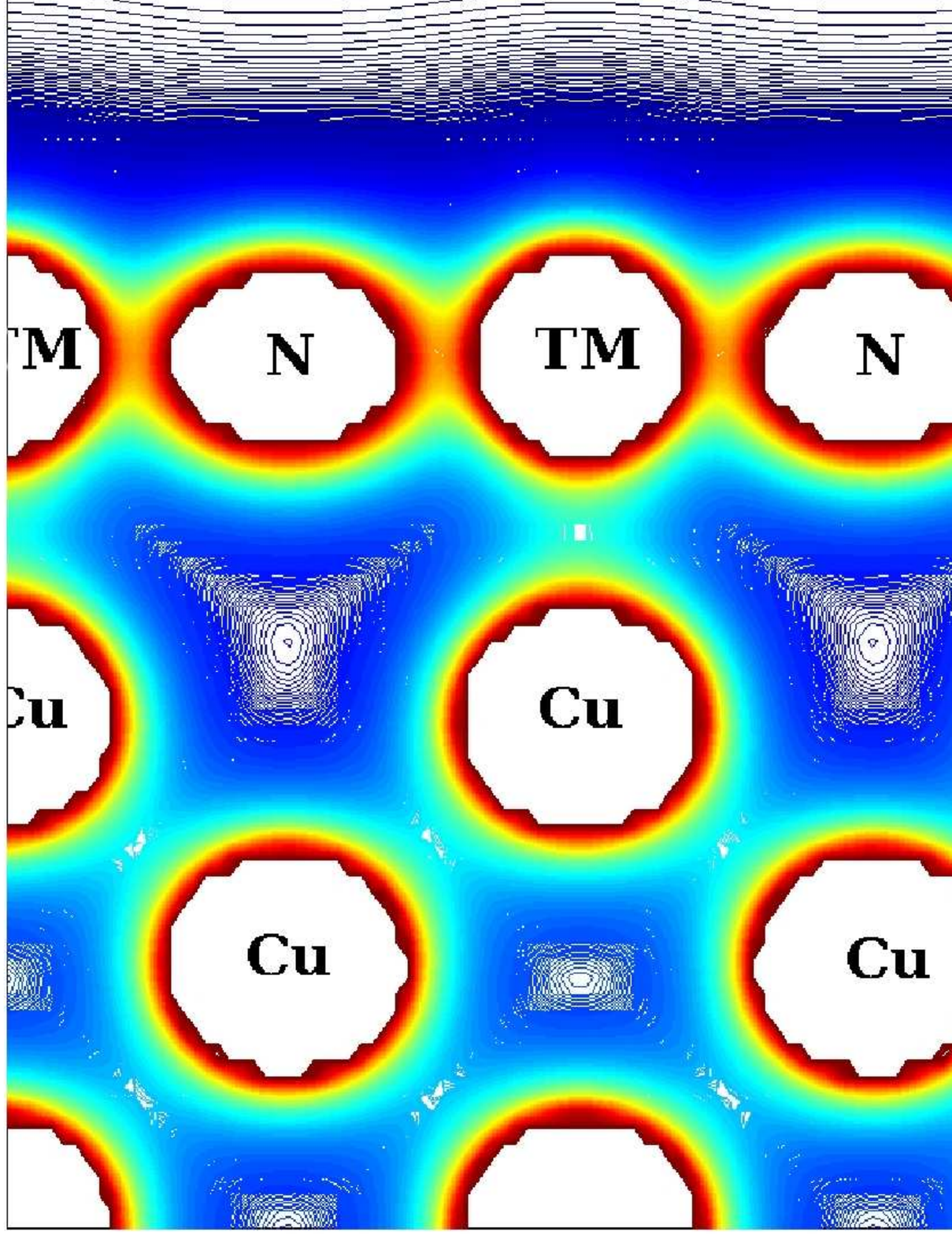,height=4.8cm,clip=} &
\psfig{file=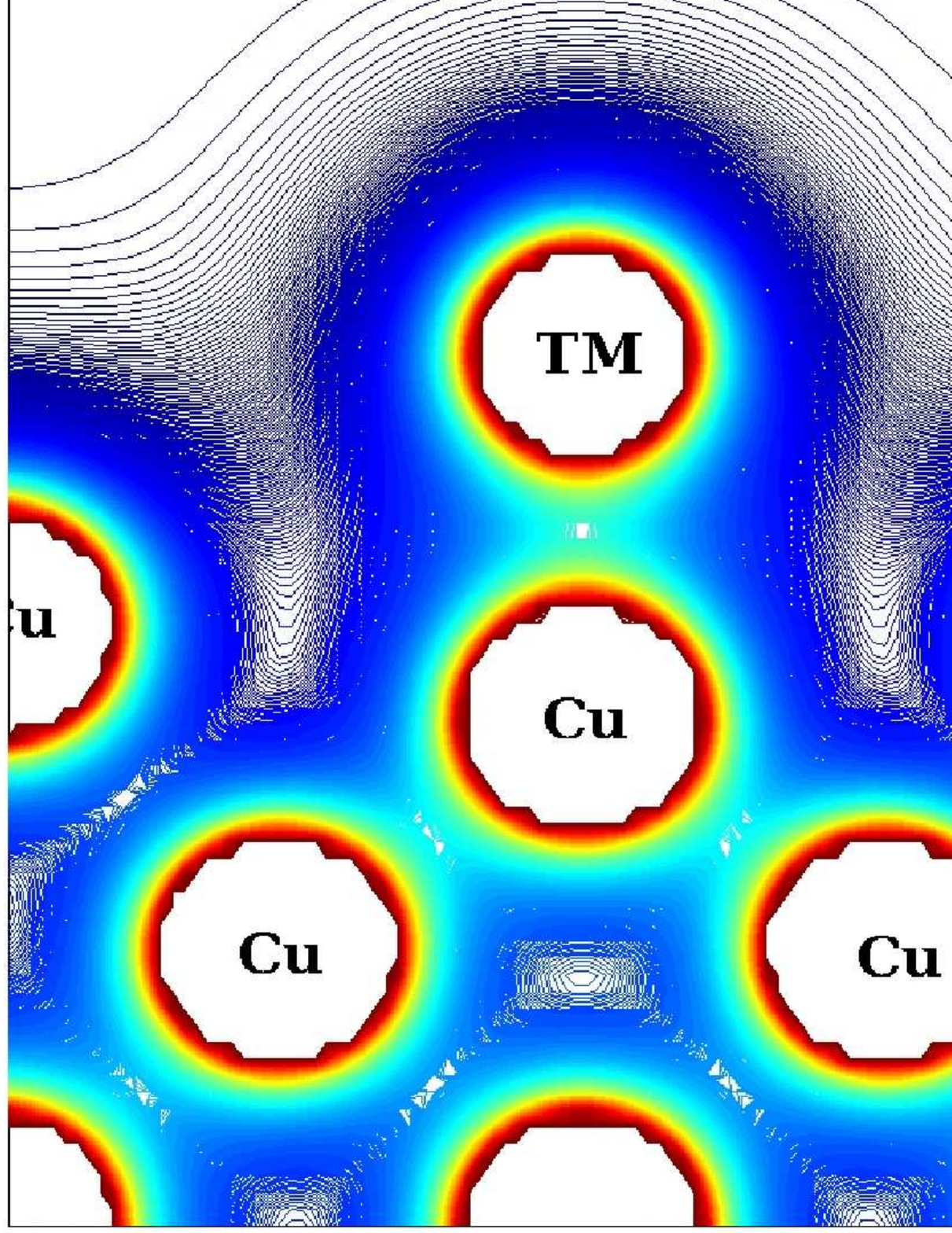,height=4.8cm,clip=} &
\psfig{file=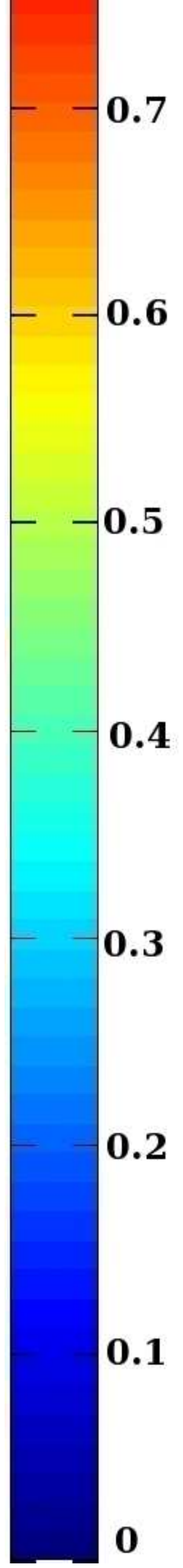,height=4.8cm,clip=}\\
\hspace{18mm}( a) & \hspace{18mm} (b) &  \\
\end{tabular}
\caption{\label{fig:fig5} (Color online) Charge density plot for a  Mn chain deposited on Cu$_2$N/Cu(001) 
in the \emph{s1} configuration, (a) along the 
chain, (b) perpendicular to the chain.}
\end{figure}

In Fig.~\ref{fig:fig5} we present electronic charge density plots for the
\emph{s1} system along planes: (a) containing the deposited chain  and (b) 
 perpendicular to it and going through a chain atom.
The \emph{s1} structure shows 1D covalent bonding   
among each 3$d$ atoms and the N ones which build the  diatomic chain.
The binding to the substrate of this covalent chain  is achieved through 
copper atoms which originally belonged to the pristine Cu$_2$N 
monolayer and that now sit below the 3$d$ atoms of the chain. This binding  
can be considered as "metallic-like" when compared with the charge
distribution inbetween Cu atoms of a Cu substrate. The bonding in this case
belongs to type i).

\begin{figure}[ht]
\centering
\begin{tabular}{p{3.8cm}p{3.8cm}p{0.6cm}}
\psfig{file=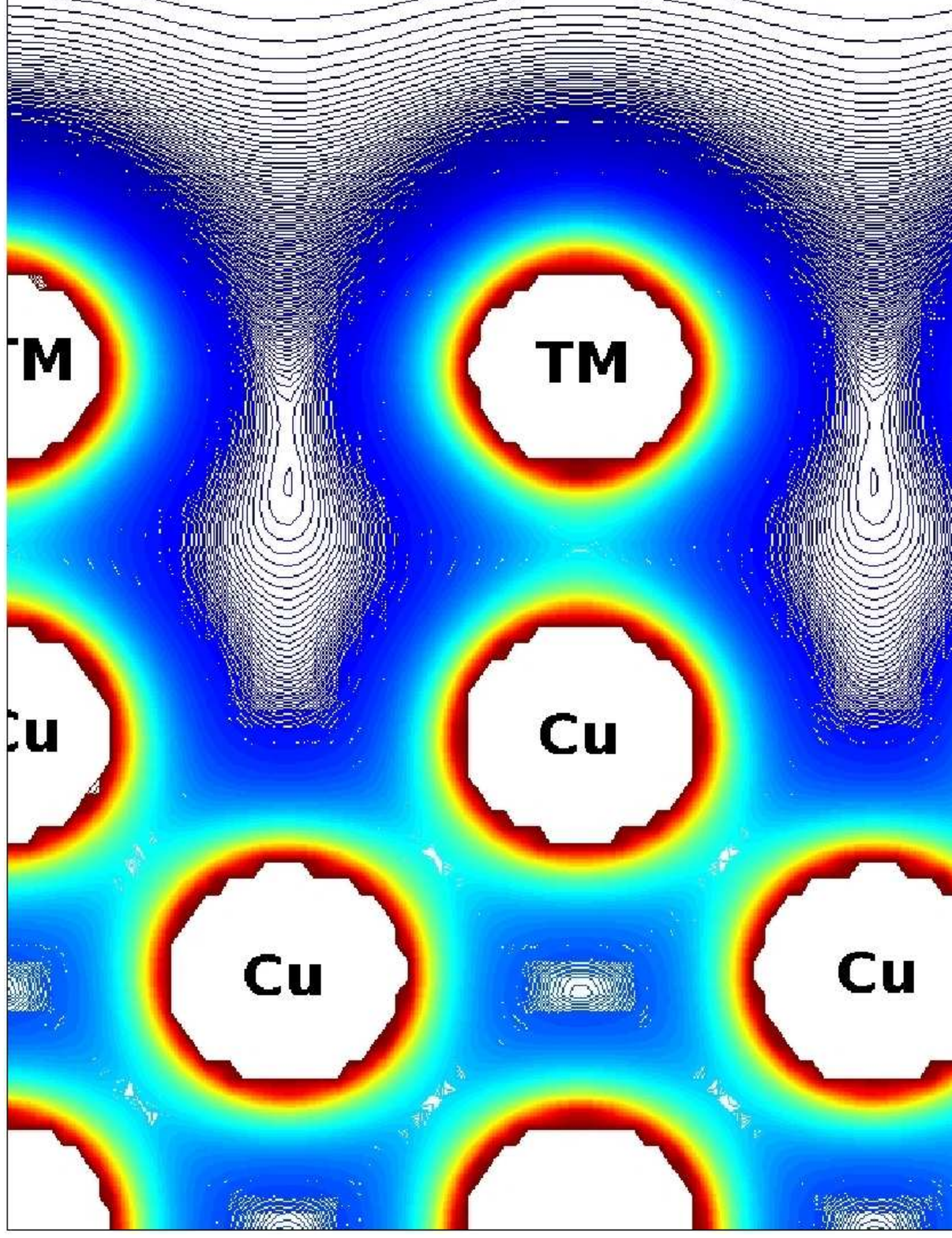,height=4.8cm,clip=} &
\psfig{file=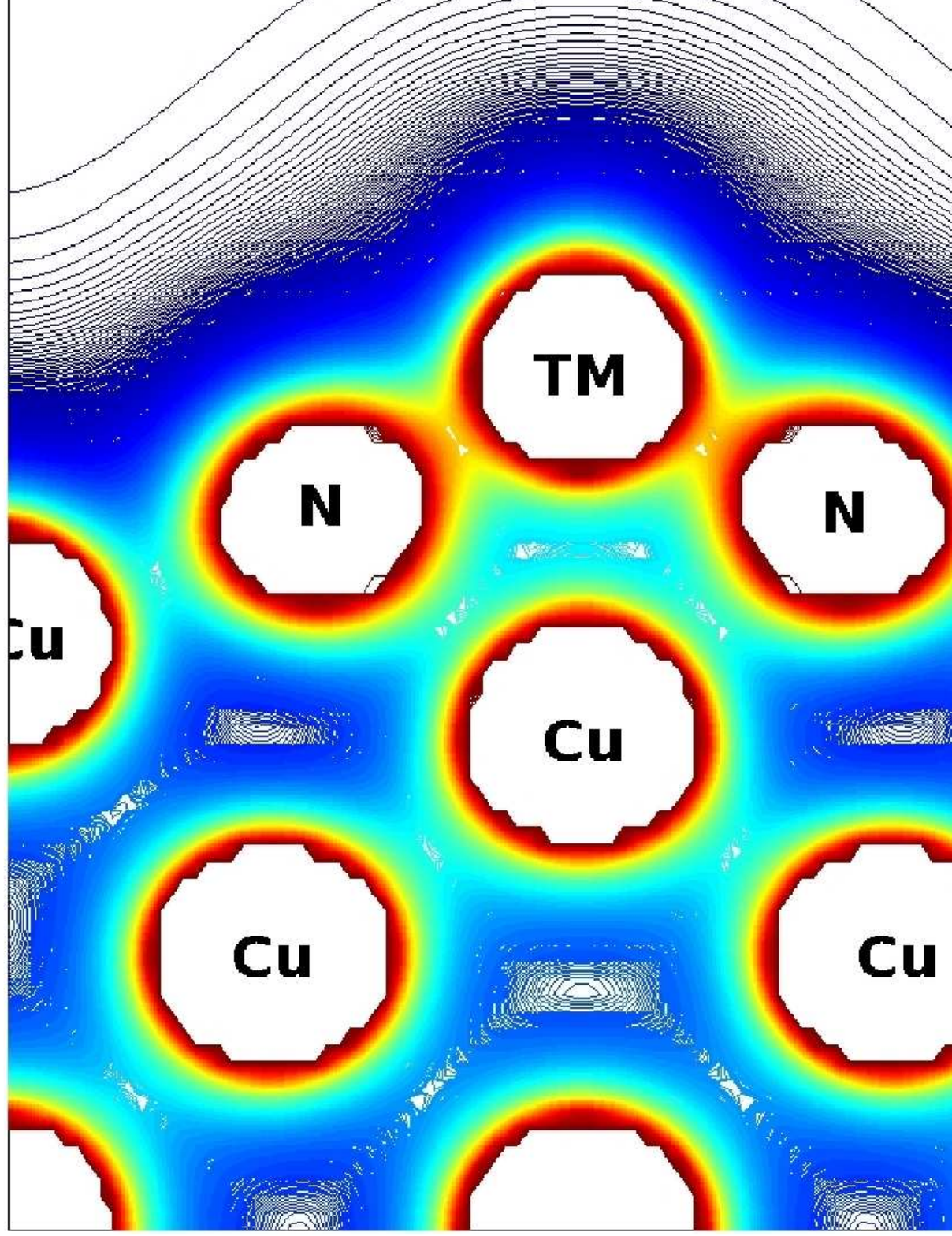,height=4.8cm,clip=} &
\psfig{file=escala.pdf,height=4.8cm,clip=}\\
\hspace{18mm}( a) & \hspace{18mm} (b) &  \\
\end{tabular}
\caption{\label{fig:fig6} (Color online) Charge density plots for a Mn chain on Cu$_2$N/Cu(001) 
in the \emph{s2} configuration (a) along
the chain, (b) perpendicular to the chain and containing a chain atom.}
\end{figure}

In the \emph{s2} configuration, the 3$d$ chains  are covalently bonded to the 
substrate through the two nitrogen atoms lying lateral to each chain one.  
These nitrogen atoms remain being covalently bonded to the original monolayer.
This can be concluded from Fig.~\ref{fig:fig6}, where electronic 
density plots for different planes are being shown. In Fig.~\ref{fig:fig6} (a)
the selected plane contains a 3$d$ chain and is perpendicular
to the surface. In Fig.~\ref{fig:fig6} (b) the plane is 
perpendicular to the chain  and shows that the N atoms on each side, 
lateral to a chain atom, form covalent bonds with it. We are then in the 
presence of a type ii) bonding.

\begin{figure}[ht]
\centering
\begin{tabular}{p{3.8cm}p{3.8cm}p{0.6cm}}
\psfig{file=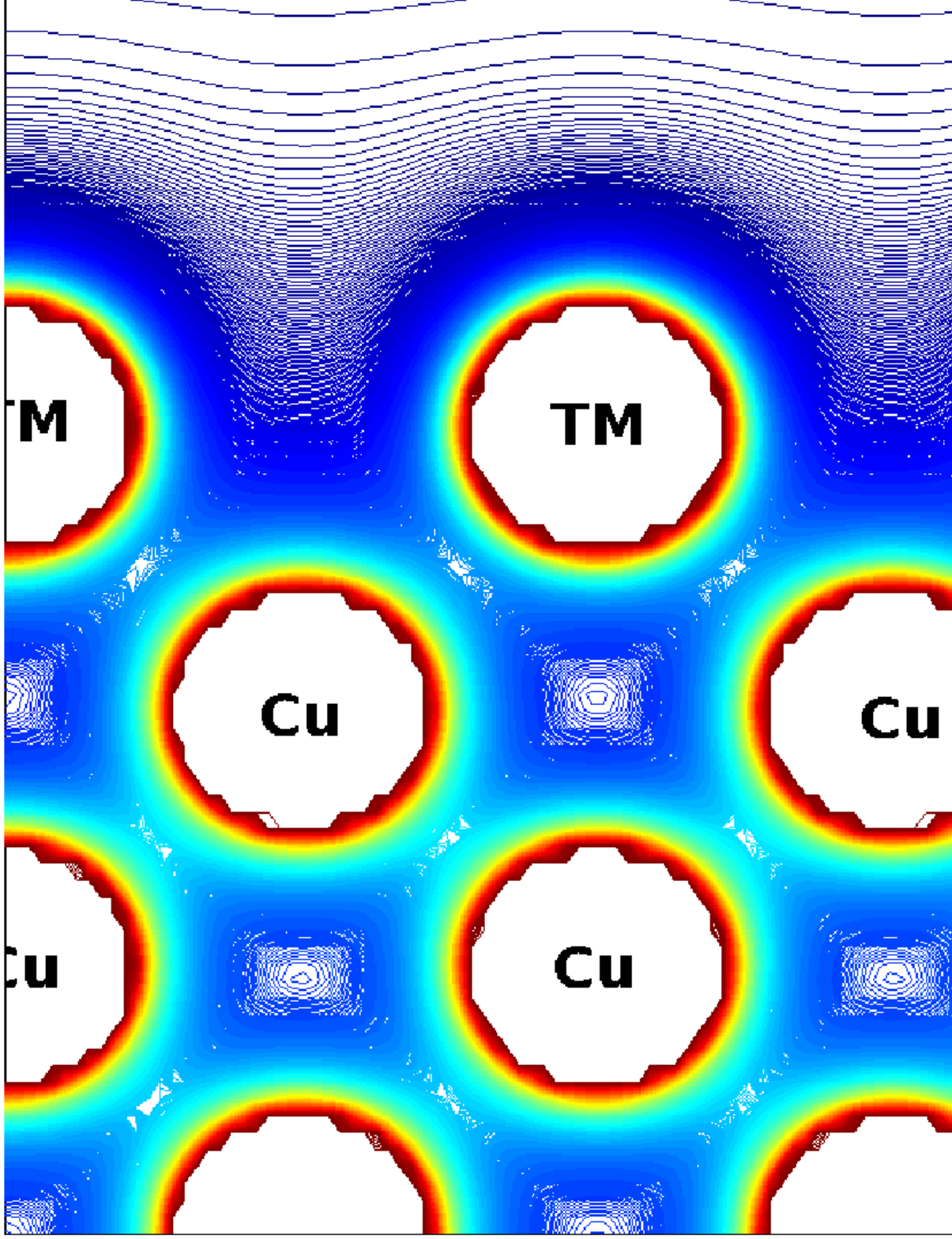,height=4.8cm,clip=} &
\psfig{file=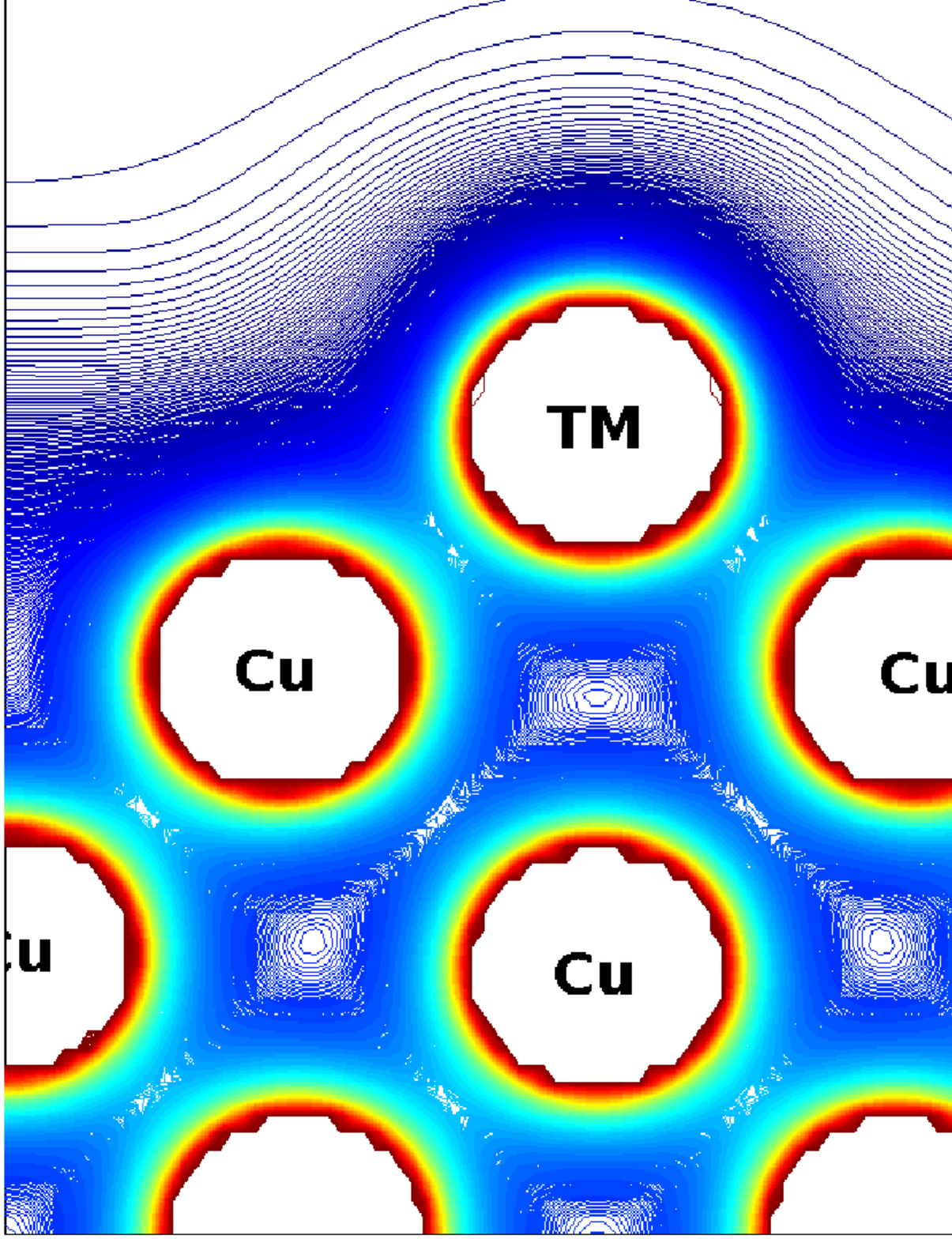,height=4.8cm,clip=} &
\psfig{file=escala.pdf,height=4.8cm,clip=}\\
\hspace{18mm}( a) & \hspace{18mm} (b) &  \\
\end{tabular}
\caption{\label{fig:fig8} (Color online) Charge density plot for a Mn chain on Cu$_2$N/Cu(001) 
in the \emph{s3} configuration, (a) along a plane paralell to the chain, (b) perpendicular 
to the chain and going through a chain atom.}
\end{figure}

\begin{figure}[ht]
\centering
\begin{tabular}{p{3.8cm}p{3.8cm}p{0.6cm}}
\psfig{file=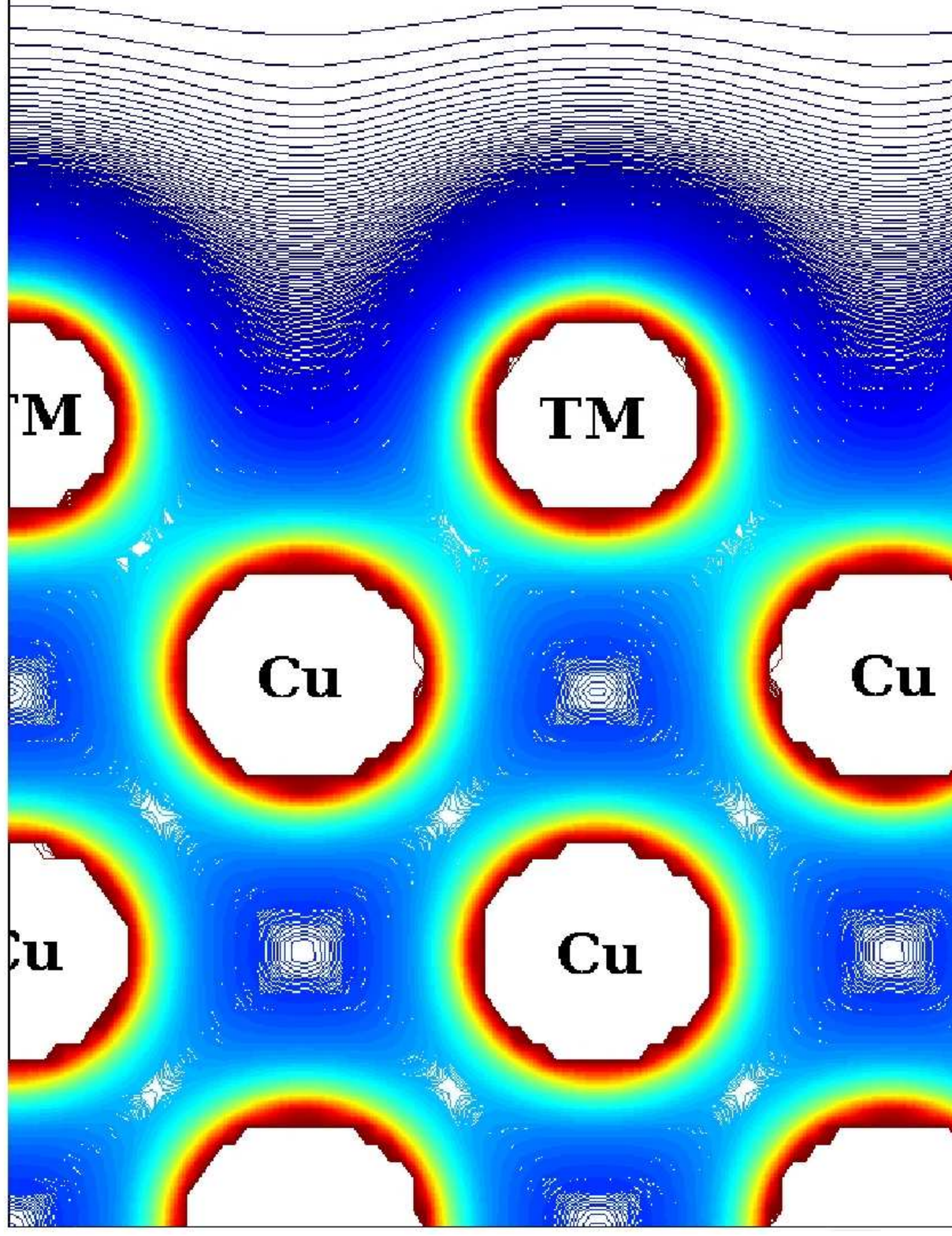,height=4.8cm,clip=} &
\psfig{file=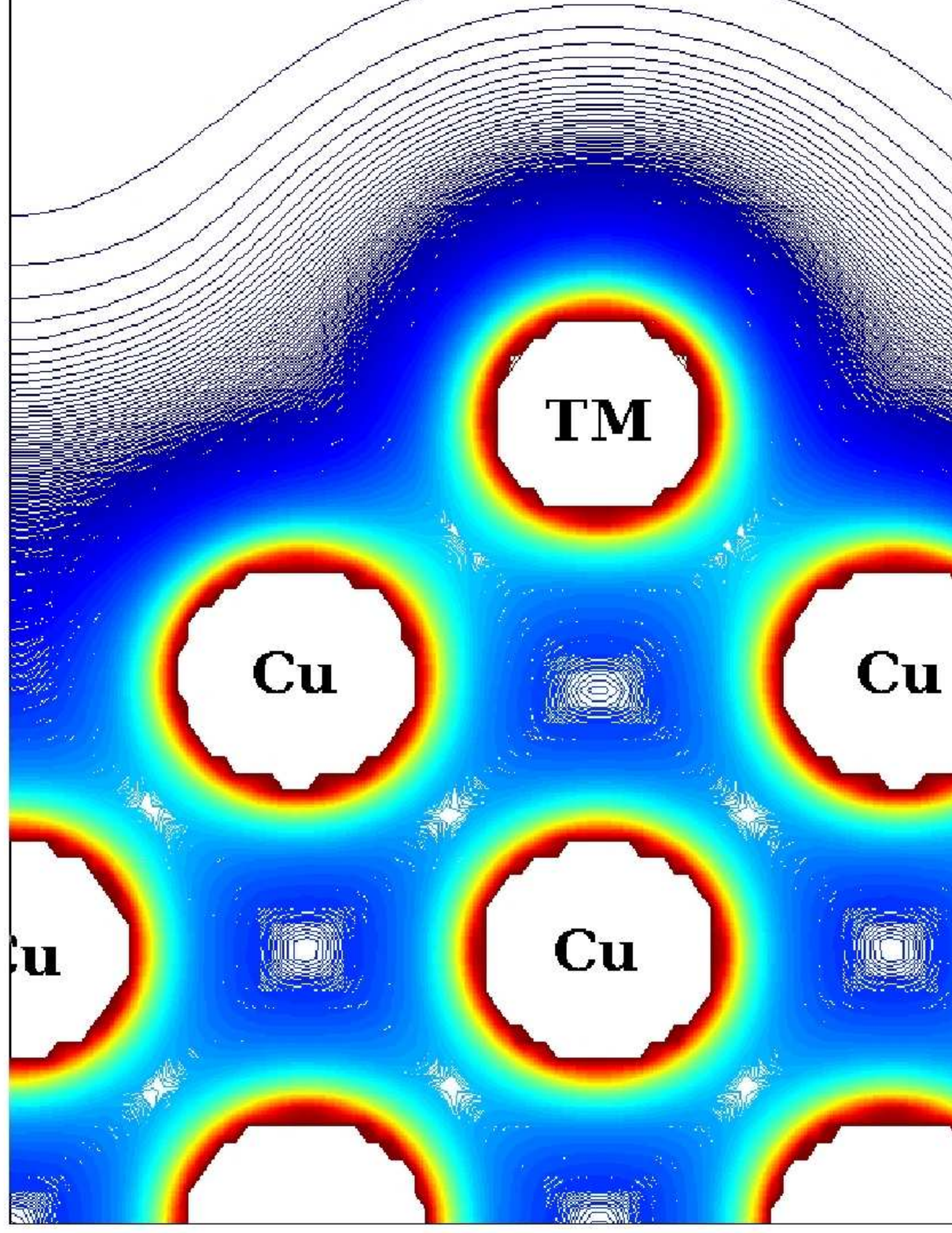,height=4.8cm,clip=} &
\psfig{file=escala.pdf,height=4.8cm,clip=}\\
\hspace{18mm}( a) & \hspace{18mm} (b) &  \\
\end{tabular}
\caption{\label{fig:fig10bis} (Color online) Charge density plot for a Mn chain on Cu(001) 
in the \emph{s5} configuration, (a) along a plane paralell to the chain (b) perpendicular 
to the chain and going through one chain atom.}
\end{figure}

In the \emph{s3} structure the adsorbed 3$d$ chain sits in a metallic-like 
channel of the surrounding Cu$_2$N monolayer, the channel being built by the
copper atoms nearest to the chain (case iii).
The bond is metallic like as it can be observed in Fig.~\ref{fig:fig8}.
Fig.~\ref{fig:fig8} (a) displays  the  electronic density along a plane which 
contains the 3$d$ chain and is perpendicular to the surface and 
Fig.~\ref{fig:fig8} (b)  shows it in a plane perpendicular
to the chain which contains one 3$d$ atom. The local environment of the 
transition metal atoms is similar to the one they have  when
adsorbed directly on Cu(001). Just for comparison, 
we present in Fig.~\ref{fig:fig10bis} the electronic density distribution  
along and perpendicular to a 3$d$ chain absorbed on holes of the Cu(001) 
surface. 
We call this last system \emph{s5} and it is clear that the environment of 
the chain in \emph{s3} and \emph{s5} is very similar.   
We have checked that there is no covalent bond between the chain
atoms and their neighboring nitrogens and this strengthens the fact that 
the chain presents only metallic bonds with its surroundings.

\begin{figure}[ht]
\centering
\begin{tabular}{p{3.8cm}p{3.8cm}p{0.6cm}}
\psfig{file=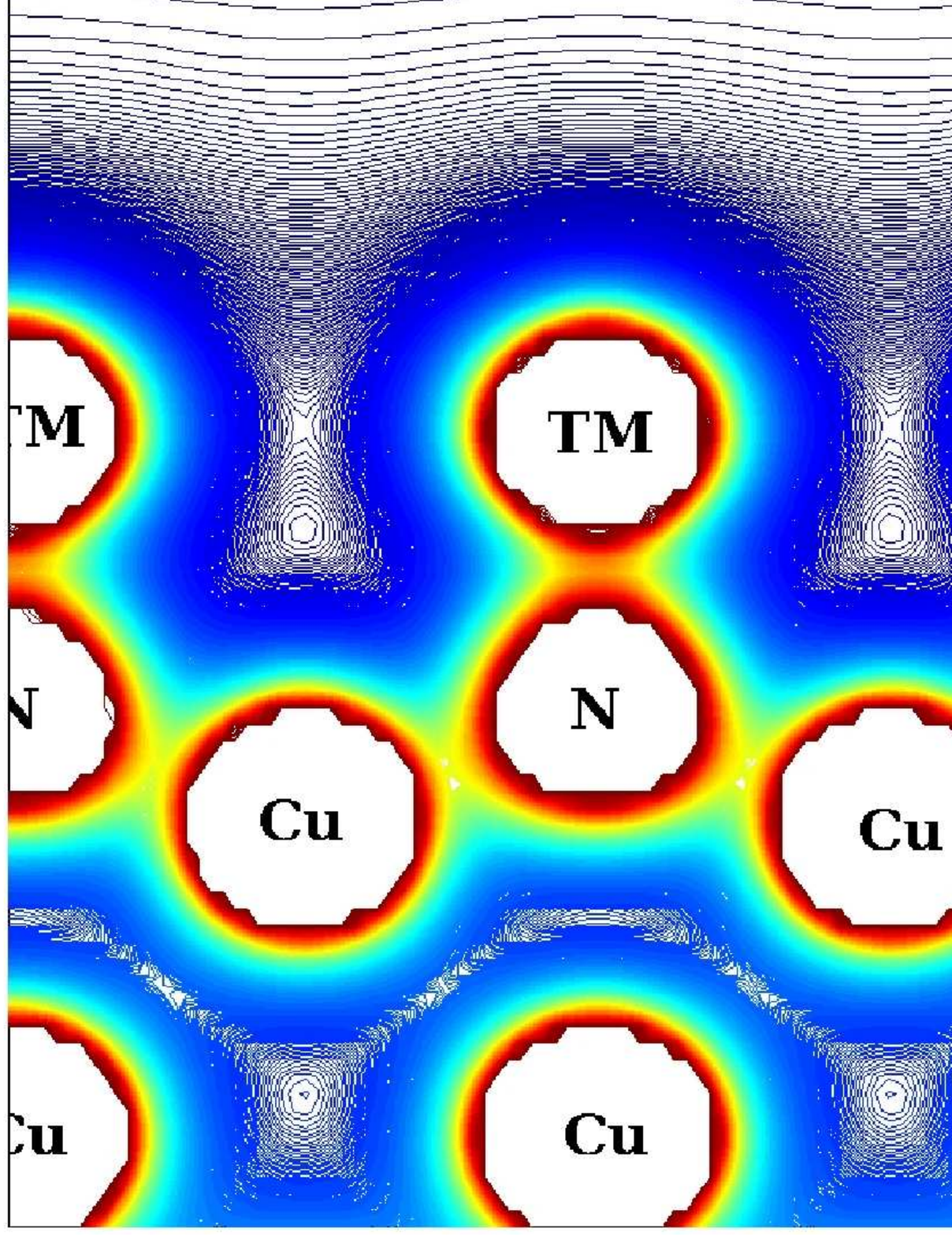,height=4.8cm,clip=} &
\psfig{file=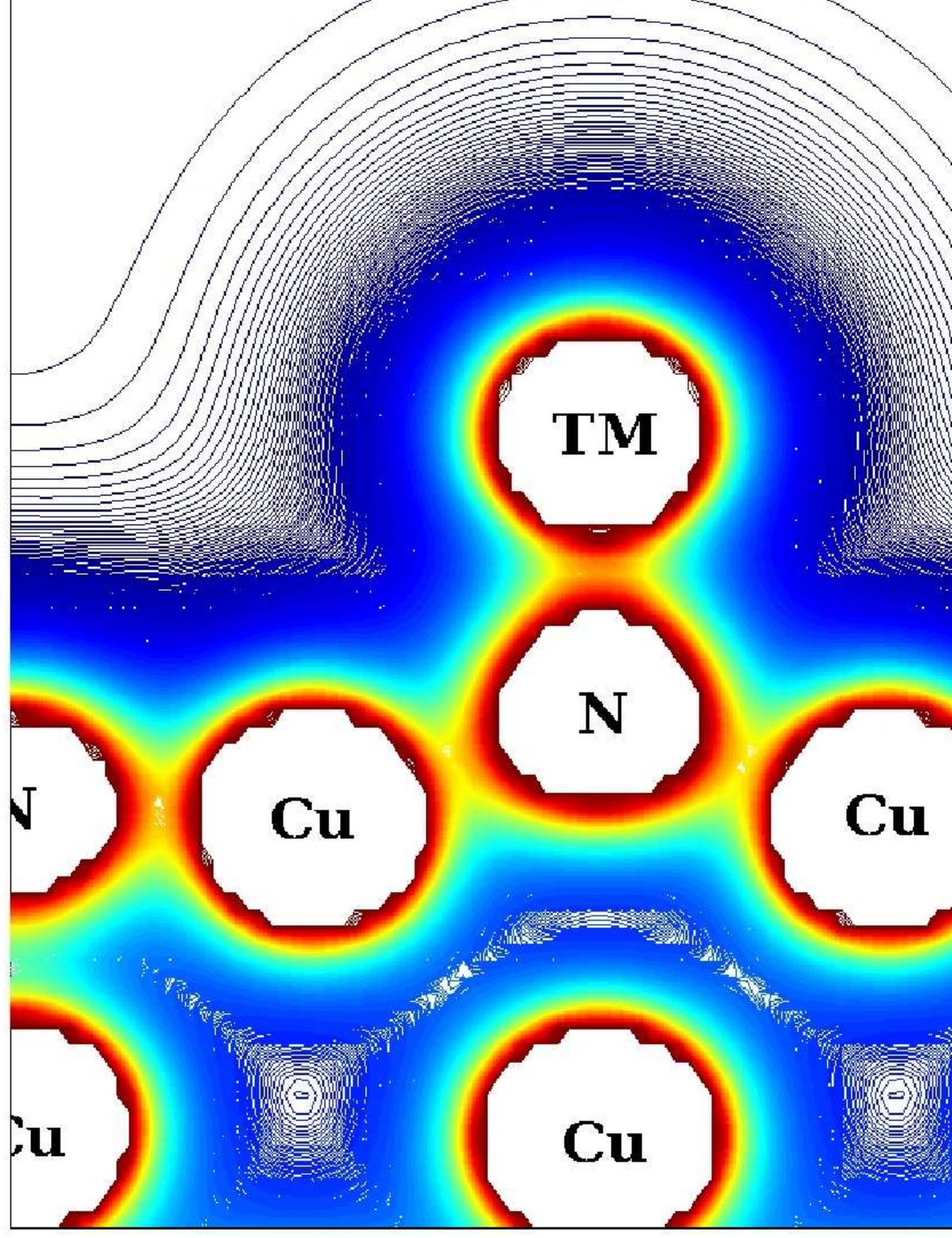,height=4.8cm,clip=} &
\psfig{file=escala.pdf,height=4.8cm,clip=}\\
\hspace{18mm}( a) & \hspace{18mm} (b) &  \\
\end{tabular}
\caption{\label{fig:fig10} (Color online) Charge density plot for a Mn chain on 
Cu$_2$N/Cu(001) in the \emph{s4} 
configuration, (a) along a plane paralell to the chain (b) perpendicular to the chain and going through one chain atom.}
\end{figure}

In the \emph{s4} configuration the 3$d$ chain is covalently bonded to 
the Cu$_2$N substrate through the nitrogen atom sitting below each chain
atom. This can be observed in the electronic density plots of 
Fig.~\ref{fig:fig10}. Fig.~\ref{fig:fig10} (a) presents a plane 
containing the 3$d$ chain, which is perpendicular to the surface 
and Fig.~\ref{fig:fig10} (b) shows a plane perpendicular
to the chain and to the surface, which goes through one atom of the chain. 
The binding in this arrangement is clearly of type ii).

All the electronic density plots shown in this work are for Mn chains,
but no relevant differences are observed for the other $d$ fillings.

Summarizing in \emph{s1}, \emph{s2} and \emph{s4} the 3$d$ atoms of the 
chains are covalently bonded to their nearest neighbor nitrogen atoms,  
but, in \emph{s1} the chain binds 
metalically to the substrate, while in \emph{s2} and \emph{s4} it does  it covalently.  
On the other hand, 
in \emph{s3} the 3$d$ chain is metalically bonded to the Cu$_2$N substrate.

\subsection{Magnetic configurations}

As stated in the introduction, among the possible magnetic solutions, we
consider only collinear ferromagnetic (FM) and antiferromagnetic (AFM) 
ones for each atomic adsorption configuration.
We are aware of the fact that in those cases in which the difference 
in energy between
the FM and AFM  magnetic order is small, other interactions could turn 
out being important and that the ground state might be non collinear,  but,  
in this work we are mainly concerned with general trends.

\begin{figure*}[ht]
\centering
\psfig{file=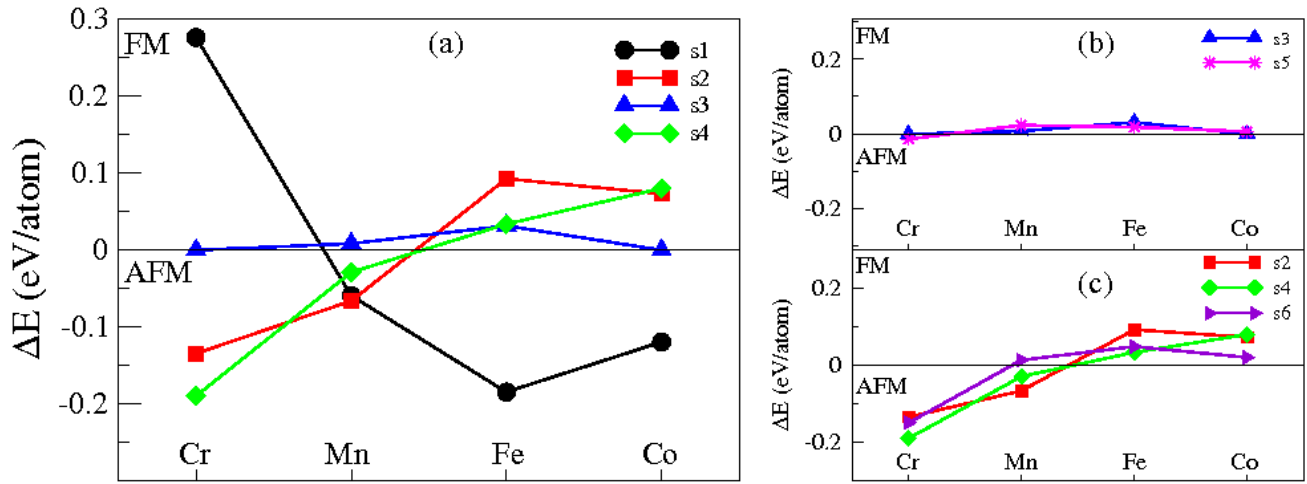,width=15cm,angle=-90,clip=i}
\caption{\label{fig:fig11}(Color online) $\Delta E$,  difference in 
energy between the AFM and FM
collinear chain  configurations, as a function of $d$ filling for
(a) the four different adsorption configurations considered, (b)  for
\emph{s3} compared with \emph{s5} (chains  adsorbed directly on Cu(001)) and, 
(c) for the \emph{s2} and  \emph{s4} configurations compared with 
\emph{s6}  (unsupported chains).}
\end{figure*}

In order to analyse the evolution of the  magnetic ground state as a 
function of $d$ band filling and 
adsorption sites of the chains, we calculate the difference 
in energy, $\Delta E$,
between the AFM and FM configurations for all the arrangements. 
If this difference
is negative the magnetic ground state is AFM, if not, it is FM.
The results obtained for \emph{s1}, \emph{s2}, \emph{s3} and \emph{s4} 
appear in Fig.~\ref{fig:fig11}.
As it can be seen \emph{s1}  has a completely different magnetic behaviour 
than the other three configurations.
In \emph{s1} the magnetic interaction goes from FM  to AFM with
increasing  $d$ filling, while the reverse is observed for  \emph{s2} 
and \emph{s4}, both, presenting similar trends.
As discussed in the previous section, in the \emph{s2} and \emph{s4} 
 arrangements the 3$d$ atoms are covalently 
bonded to neighboring nitrogens, but in \emph{s1} there is a nitrogen atom 
inbetween two 3$d$ ones.  The nitrogen atom in the diatomic chain gives
rise to a superexchange like magnetic coupling among the  3$d$ atoms.
This kind of adsorption situation, which enables  this nitrogen 
mediated interaction, 
has already been treated in the literature for Mn chains and for 
several dimers.\cite{Jones_Ti,Jones_Lin,Scopel}

In \emph{s2} and  \emph{s4} the atoms of
the 3$d$ chains are covalently bonded  to the substrate through
their neighboring nitrogen atoms and in both cases the magnetic ground
states go from AFM to FM as a function of $d$ 
filling. There is a direct exchange interaction along the chains and an 
indirect interaction through the substrate. The question is which is
more significant. 
If one compares the results for $\Delta E$ obtained in the case 
of the \emph{s2} and  \emph{s4} 
configurations with those corresponding to the unsupported chains at 
the same interatomic distance, \emph{s6},
the same trends for the magnetic interaction are found as a function 
of $d$ filling.
The energy differences  between ferro- and antiferromagnetic configurations 
are also of the same order of magnitude.  
The results for $\Delta E$ for the unsupported chains together with those 
corresponding to \emph{s2} and  \emph{s4} 
are presented in Fig~\ref{fig:fig11} (c). These results show
that the direct interaction among 
the 3$d$ atoms is comparable in the three cases.  
This is also apparent from the comparison of the electronic density 
plots of Fig~\ref{fig:fig6} (a) and
Fig~\ref{fig:fig10} (a) with that of an unsupported chain at the same 
interatomic distance.
The indirect interaction through the substrate is certainly affecting 
the energy difference for each particular filling, giving rise to a 
switching of the solution from FM to AFM in the case of Mn. In the case 
of Mn the direct interaction is small and competes with the 
other interactions at the considered interatomic distance, but it does 
not change the general trends.

Finally, \emph{s3}  presents a different behavior as a function of $d$ 
filling when compared with the other configurations and the results for 
$\Delta E$ point towards the preponderance of another type of magnetic 
interaction.
The difference in energy $\Delta E$, in this case, evolves  as if the chains were adsorbed directly
on hollow sites of Cu(001), that is, the so called  \emph{s5} configuration.
This is clear from  Fig~\ref{fig:fig11} (b)  where we compare $\Delta E$ for
\emph{s3} and \emph{s5} and from which it is seen that the order of 
magnitude and the sign of $\Delta E$ is the same for the four $d$ filling
studied. In the \emph{s3} arrangement the 
interaction among atoms of the 3$d$ chains
seems to be predominantly  RKKY-like,  as it is the case when 3$d$ atoms are 
directly adsorbed on Cu(111).\cite{Wiensendanger} 

It is interesting to compare our results with those obtained for the
exchange coupling among Ti atoms in Ref.~\onlinecite{Jones_Ti} when depositing  a
monolayer of titanium  on the same Cu2N/Cu(001)surface.  The coupling
obtained by the authors along an N axis and along a hollow axis are to
be compared with  the exchange energies obtained by us in the \emph{s1} 
and \emph{s2} configurations of the deposited chains.  Following the trends
already discussed, for Ti chains in the \emph{s1} and \emph{s2} 
configurations we would have obtained also ferromagnetic and 
antiferromagnetic exchange couplings, respectively.

We have repeated the calculations within the DFT+U framework\cite{Anisimov,Liech} 
using $U=5$ eV for Cr and Mn
and $U=2$ for Fe and Co. The results obtained for $\Delta E$ do no change the 
general trends obtained.

\section{Conclusions}

Summarizing,  the covalently bonded nitride substrate Cu$_2$N  offers an 
interesting template  
to taylor magnetic interactions among transition metal atoms by taking as variable the adsorption sites.
We have shown this by depositing $3d$  chains  on four different adsorption 
sites with the same interatomic distance among 3$d$ atoms and for four different $d$ fillings. 
Due to the different type of bonding present, we obtain
that the $3d$ atoms  interact magnetically through completely different 
mechanisms depending on adsorption geometry. 
For a given $d$ filling, the interaction  can be strongly  AFM or  FM for the same interatomic 
distance along the chain. This versatility could be  very promising 
regarding the future possibility of 
making spin engineering at the nanometer scale. 
We expect a similar richness in the magnetic interactions, when using other covalently  
bonded substrates.

\section*{Acknowledgments}

MAB is supported and MCU and  AML are partially supported by CONICET
and belong to the Institute if nanoscience and Nanotechnology
(INN) of the Atomic Energy Agency (CNEA), Argentina.  This work was
partially funded  by the grants PIP No 11220090100258  (CONICET),
UBACYT-X123 (UBA), PICT R1776 (ANPCyT) and PRH 2077 No 74 (ANPCyT).

\end{document}